\documentclass[aps,prx,twocolumn,superscriptaddress,floatfix]{revtex4-1}
 
\usepackage{graphicx}
\usepackage{dcolumn}
\usepackage{verbatim}
\usepackage[colorlinks=true]{hyperref}
\usepackage{multirow}
\usepackage{bm}
\usepackage{times}
\usepackage{color}
\usepackage{amstext}
\usepackage{amsmath}
\usepackage{amssymb} 
\usepackage{amsfonts}
\usepackage{natbib}
\usepackage[final]{microtype}
\usepackage{csquotes,lmodern}
\usepackage{hyperref}
\hypersetup{hidelinks}

\newcommand{\TNSe}{Ta$_2$NiSe$_5$}  
\newcommand{\TNS}{Ta$_2$NiS$_5$}

\begin{document} 

\title{Giant exciton Fano resonance in quasi-one-dimensional Ta$_2$NiSe$_5$}

\author{T.~I.~Larkin}
\email{T.Larkin@fkf.mpg.de}
\affiliation{
Max Planck Institute for Solid State Research, Heisenbergstra{\ss}e~1,
70569 Stuttgart, Germany}

\author{A.~N.~Yaresko}
\affiliation{
Max Planck Institute for Solid State Research, Heisenbergstra{\ss}e~1,
70569 Stuttgart, Germany}

\author{D.~Pr\"opper}
\affiliation{
Max Planck Institute for Solid State Research, Heisenbergstra{\ss}e~1,
70569 Stuttgart, Germany}

\author{K.~A.~Kikoin}
\thanks{Deceased 21 November 2016}
\affiliation{
School of Physics and Astronomy, Tel Aviv University, 69978 Tel Aviv,
Israel}

\author{Y.~F.~Lu}

\affiliation{
Department of Physics, The University of Tokyo, Hongo, Tokyo 113-0033,
Japan}

\author{T.~Takayama}
\affiliation{
Max Planck Institute for Solid State Research, Heisenbergstra{\ss}e~1,
70569 Stuttgart, Germany}

\author{Y.~-~L.~Mathis}
\affiliation{
Synchrotron Facility ANKA, Karlsruhe Institute of Technology, 76344
Eggenstein - Leopoldshafen, Germany}

\author{A.~W.~Rost}
\affiliation{
Max Planck Institute for Solid State Research, Heisenbergstra{\ss}e~1,
70569 Stuttgart, Germany}
\affiliation{
Institute for Functional Materials and Quantum Technology, University
of Stuttgart, Pfaffenwaldring 57, 70550 Stuttgart, Germany}

\author{H.~Takagi}
\affiliation{
Max Planck Institute for Solid State Research, Heisenbergstra{\ss}e~1,
70569 Stuttgart, Germany}
\affiliation{
Department of Physics, The University of Tokyo, Hongo, Tokyo 113-0033,
Japan}
\affiliation{
Institute for Functional Materials and Quantum Technology, University
of Stuttgart, Pfaffenwaldring 57, 70550 Stuttgart, Germany}

\author{B.~Keimer}
\affiliation{
Max Planck Institute for Solid State Research, Heisenbergstra{\ss}e~1,
70569 Stuttgart, Germany}

\author{A.~V.~Boris} 
\email{A.Boris@fkf.mpg.de}
\affiliation{
Max Planck Institute for Solid State Research, Heisenbergstra{\ss}e~1,
70569 Stuttgart, Germany}


\begin{abstract}
We report the complex dielectric function of the quasi-one-dimensional chalcogenide
\TNSe, which exhibits a structural phase transition that has been attributed
to exciton condensation below $T_c = 326$ K \cite{Wakisaka2009,Lu2017},
and of the isostructural \TNS{} which does not exhibit such a transition.
Using spectroscopic ellipsometry, we have detected exciton doublets with
pronounced Fano lineshapes in both the compounds. The exciton Fano resonances
in \TNSe{} display an order of magnitude higher intensity than those in \TNS.
In conjunction with prior theoretical work \cite{Rashba1975}, we attribute
this observation  to the giant oscillator strength of spatially extended
exciton-phonon bound states in \TNSe. The formation of exciton-phonon complexes
in \TNS{} and \TNSe{} is confirmed by the pronounced temperature dependence
of sharp interband transitions in the optical spectra,  whose peak energies
and widths scale with the thermal population of optical phonon modes. The
description of the optically excited states in terms of strongly overlapping
exciton complexes is in good agreement with the hypothesis of an EI ground
state. 
\end{abstract}

\maketitle
\section{Introduction}
Cooper pairing of fermions is one of the most fundamental and successful
concepts of condensed matter physics. In analogy to the superconducting state
generated by pairing of electrons in metals, condensation of neutral electron-hole
pairs in semimetals or narrow-band semiconductors has been predicted to generate
an \textquotedblleft excitonic insulator\textquotedblright \ (EI) state,
which has so far remained elusive \cite{Keldysh1965,Jerome1967,Rice1968,Kunes2015}.
Bose condensation of excitons has been experimentally explored on various platforms, including nonequilibrium excitons in optically pumped semiconductors, indirect excitons in coupled quantum wells, and highly excited exciton-polaritons in semiconductor microcavities \cite{Ivanov2007}. In contrast to these systems, the EI scenario predicts spontaneous coherence between conduction and valence bands in a bulk material in thermal equilibrium, if the exciton binding energy exceeds the band gap. 

In the absence of distinct manifestations of macroscopic quantum coherence (such as dissipationless currents and the Meissner effect in superconductors), the experimental identification of an EI state is a formidable challenge. EI instabilities have been proposed to occur in the chalcogenide compounds TmSe$_{1-x}$Te$_{x}$ and Sm$_{1-x}$La$_{x}$S when their indirect band gaps are closed under external pressure~\cite{Wachter1991,Wachter2004,Wachter2013,Bronold2006,Wachter1995,Kikoin1983,Kikoin1984}. In $1T$--TiSe$_2$, a distinct flattening of the valence band observed by angle-resolved photoemission spectroscopy (ARPES) has been interpreted as a manifestation of the EI transition~\cite{Aebi2007,Aebi2009,Aebi2010}. However, the theoretical description of the electron system in $1T$--TiSe$_2$ is confounded by a charge density wave instability \cite{NLWang2007,Kidd2002,Zenker2013,Porer2014,Aebi2015,Watanabe2015}. Moreover, distinct excitonic absorption features have not yet been observed in spectroscopic experiments on any of the EI candidates, perhaps because the indirect nature of the band gaps implies that the formation of excitons is necessarily accompanied by phonons \cite{Phan2013,Zenker2014}.

More recently, the EI was proposed as the ground state of the direct-gap semiconductor \TNSe{} \cite{Wakisaka2009,Lu2017}, which is built up of parallel sets of Ni and Ta chains (Fig.~1(a)) and thus exhibits a quasi-one-dimensional electronic structure with a high joint density of states for electron-hole pair excitations. The valence and conduction bands are composed of hybridized Ni $3d$ -- Se $4p$ levels and Ta $5d$ levels, respectively. The spatial separation of holes in the Ni chains and electrons in the Ta chains is expected to reduce the probability for electron-hole recombination across the direct band gap, and to enhance the exciton lifetime. At a critical temperature $T_c = 326$ K, \TNSe{} undergoes a second-order transition from a high-temperature orthorhombic structure  to a low-temperature structure with a subtle monoclinic distortion and reduced electrical conductivity \cite{Sunshine1985,Salvo1986}. ARPES experiments motivated by the favorable features of the electronic structure of \TNSe{} revealed a characteristic flattening of the bands upon cooling below the phase transition \cite{Wakisaka2009,Wakisaka2012,Seki2014}, akin to the phenomenology of $1T$--TiSe$_2$ but without complications arising from CDW formation \cite{Kaneko2013,Kaneko2014}. Despite this promising situation, direct evidence of excitons and their behavior across the purported EI transition has thus far not been reported.

Motivated by the desire to explore the exciton states directly, we have used wide-band spectroscopic ellipsometry to accurately determine the optical conductivity and permittivity of  \TNSe{} and the isostructural compound \TNS, which does not exhibit a phase transition~\cite{Sunshine1985,Salvo1986} and thus serves as a reference. The chain structure of both compounds gives rise to highly anisotropic electrodynamics. 
For photon polarization along the chains, exciton Fano resonances were unambiguously identified in both compounds at low temperatures. The oscillator strengths of the exciton Fano resonances in \TNSe{} are found to be of giant magnitude, an order of magnitude larger than those in \TNS. We will discuss the origin of the large spectral weight in \TNSe{} and its relationship to the EI hypothesis.

\section{Methods}

We performed direct ellipsometric measurements of the complex dielectric
function, $\varepsilon (\omega) = \varepsilon_1 (\omega)+i\varepsilon_2 (\omega)=1+4\pi
i[\sigma_1 (\omega)+i\sigma_2 (\omega)]/\omega$, over a range of photon energies
extending from the far infrared ($\hbar\omega = 0.01$ eV) into the ultraviolet
($\hbar \omega = 6.5$ eV). The $a$-axis component of $\varepsilon (\omega)$
corresponds to the measured pseudodielectric function $\langle \varepsilon_a\rangle
\approx \varepsilon_a$ at angle of incidence ranging from 70$^{\circ}$ to
80$^{\circ}$ for sample orientations with the $a$ axis in the plane of
incidence (Appendix A).

Thin single crystals of \TNS{} and \TNSe{} with typical lateral dimensions
of $10 \times 1 \mathrm{mm}^2$ along the $a$- and $c$-axes, respectively,
were well characterized by X-ray, dc transport and specific heat measurements
\cite{Lu2017}, and cleaved before every optical measurement. In the frequency
range 10\,meV to 1\,eV we used home-built ellipsometers in combination with
Bruker IFS 66v/S and Vertex 80v FT-IR spectrometers. Some of the experiments
were performed at the IR1 beam line of the ANKA synchrotron light source
at the Karlsruhe Institute of Technology, Germany. The measurements in the
frequency range $0.6 - 6.5\,\text{eV}$ were performed with a  Woollam variable
angle ellipsometer (VASE) of rotating-analyzer type.

\begin{figure}
\includegraphics[width=.5\textwidth]{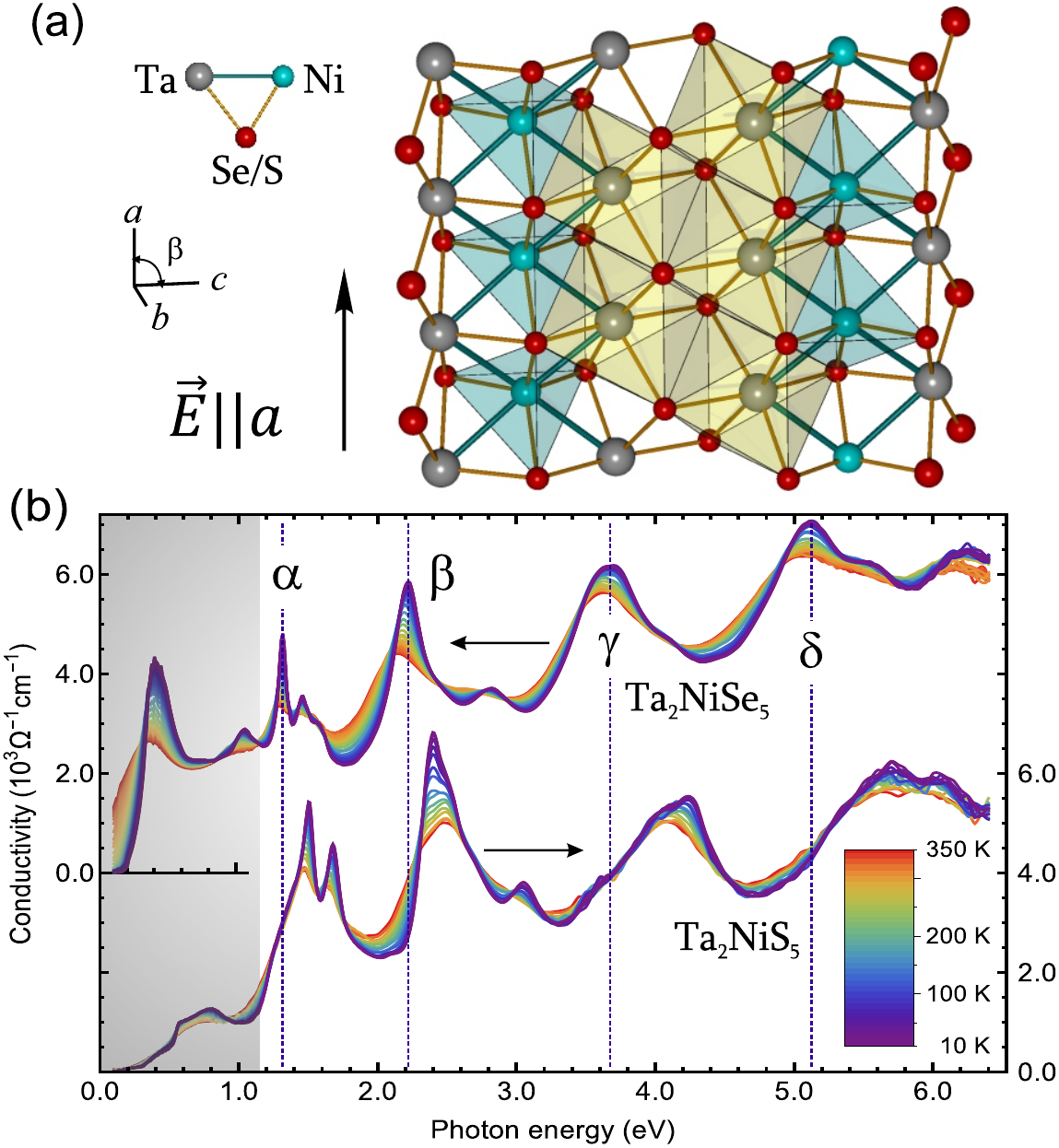}
\caption{
(a) View of a single layer of the \TNSe{} structure down the [010] direction
(not different from  \TNS{}). (b) Real part of the optical conductivity $\sigma_1(\omega)$
of \TNSe{} (upper curves, left axis) and \TNS{} (lower curves, right axis)
measured at different temperatures for incident light polarized along the
$a$ axis.}
\protect\label{fig1}
\end{figure}

\begin{figure}
\includegraphics[width=.45\textwidth]{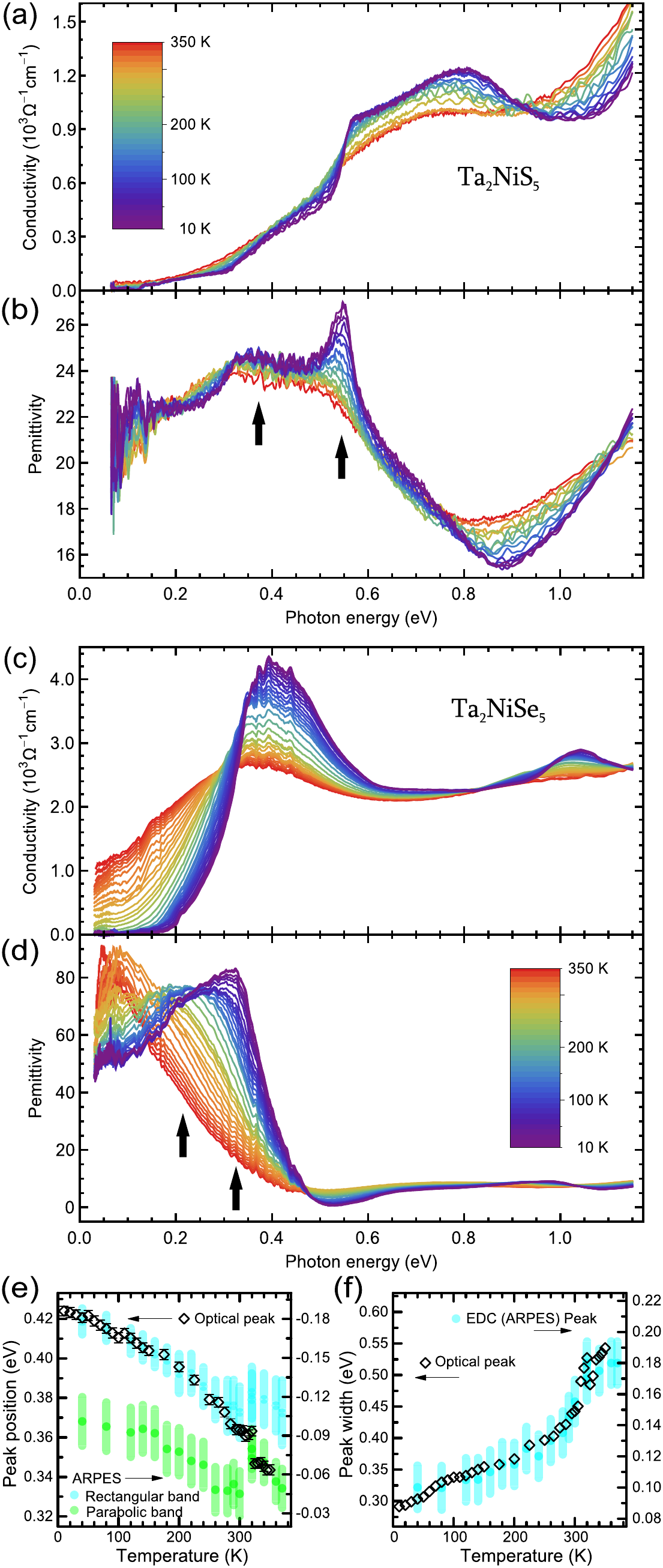}
\caption{
Real part of the infrared conductivity $\sigma_1(\omega)$ (a) and dielectric
permittivity $\varepsilon_1(\omega)$ (b) of \TNS{} measured at different
temperatures. (c) and (d) are the same as (a) and (b) but for \TNSe{}. The
arrows mark the Fano resonance peak positions.
(e) Red shift of the lowest-energy absorption band in \TNSe{} (black diamonds,
this work), along with the $T$-dependence of the peak position in the EDC
(blue bars, from Ref. \cite{Wakisaka2012}), and the vertex of the corresponding
parabolic band (green bars, from Ref.\cite{Wakisaka2012}), (f) broadening
of the optical band (black diamonds, this work) and EDC peak width (blue
bars, from Ref.\cite{Wakisaka2012}).
}
\protect\label{fig2}
\end{figure}

Relativistic LDA band structure calculations were performed for the experimental
crystal structure of \TNS{} and \TNSe{} \cite{Sunshine1985} using the linear
muffin-tin orbital method \cite{Yaresko2004}.

\section{Results and discussion}
Figure 1(b) shows the optical conductivity spectra, $\sigma_1(\omega)$, of \TNSe{} and \TNS{} for incident light polarized along the chains. The conductivity between 1.0 and 5.5 eV is dominated by a sequence of sharp peaks, which fall into distinct groups labeled $\alpha$ through $\delta$. 
Apart from shifts to higher photon energies and differences in the fine structure of the $\alpha$ bands, all features in $\sigma_1(\omega)$ of \TNS{} are in one-to-one correspondence to those in the \TNSe{} spectra. This observation agrees with density functional theory (DFT) calculations which yield closely similar band structures of both compounds; note that the monoclinic distortion in the low-temperature crystal structure of \TNSe{} does not significantly affect the band dispersions (Appendix B). The shorter Ta$-$S bonds and stronger Ta $d$ -- S $p$ hybridization, compared to Ta$-$Se, result in an increased energy separation between the Ta $d$  and S $p$ -- Ni $d$ states, in analogy to the titanium dichalcogenide compounds $1T$--TiX$_2$ (X = S, Se, Te) \cite{Reshak2003}. The peak positions of the corresponding interband transitions also move accordingly, in agreement with the optical data in Fig.~1(b).

The DFT calculations reproduce the general trend of the material dependence of $\sigma_1(\omega)$, but do not capture the main features of the dielectric response in the infrared spectral range below 1 eV. In particular, the calculations predict a metallic nonmagnetic solution for both compounds. The top of the valence band, formed by almost fully occupied S 3$p$ (Se 4$p$) and Ni 3$d$ states, overlaps with the bottom of the conduction band which is derived from hybridized Ta 5$d$ and chalcogen $p$ orbitals. The formal valencies are close to Ta$^{5+}$ (5$d^{0}$), Ni$^{0}$ (3$d^{10}$), and S(Se)$^{2-}$ ($p^{6}$). As the Ni $d$-shell is completely filled and the Ta one is empty, accounting for on-site Coulomb repulsion within the LDA+U band-structure approach does not open a gap, contrary to our infrared data which reveal clear gaps in the optical conductivity of \TNSe{} and \TNS{} at low temperature (Fig.~2(a,c)). Based on the nearly linear rise of $\sigma_1(\omega)$, we estimate the size of the band gaps as $\sim 0.16$  eV and $\sim 0.25$ eV for $T \sim150$ K in \TNSe{} and \TNS{}, respectively. For the selenide, half of this value agrees with the activation energy of $\sim $ 1000 K ($\sim $85 meV) inferred from electrical resistivity measurements \cite{Seki2014}, implying that the width of the mobility gap coincides with the width of the optical gap, and that the Fermi level lies in the middle of the gap. 
The optical gap of  \TNSe{} decreases gradually with increasing temperature and closes as $T \rightarrow T_c$ \cite{Lu2017}. 

Figures 2(e,f) show the parameters characterizing the lowest-energy absorption resonance of \TNSe{} as a function of temperature, together with ARPES data for the valence band by Wakisaka {\it et al}\cite{Wakisaka2012}.
The temperature evolution of the optical peak energy follows the shift of
the top of the valence band towards the Fermi level. The maximum energy of
the band is determined by the maximum of the ARPES energy distribution curve
(EDC) at low temperatures. Above the structural transition temperature, the
vertex of the parabolic fit of the band is better representative. The width
of the infrared band rapidly increases near the structural transition, mirroring
the anomalous broadening of the EDC peak. The high correlation of these datasets
confirms that the flat Ni $3d$ --  Se $4p$ valence band is the initial state
for its associated interband transition. The flattening and shift of the valence band have been partly addressed using finite temperature variational cluster approximation calculations on an extended Falicov-Kimball model - a minimal lattice model to describe the excitonic insulator state \cite{Seki2014}. Our data yield a complementary observation that the marked temperature-induced red shift and broadening of the corresponding infrared band with increasing temperature are closely analogous to the manifestations of small polarons in optical absorption spectra \cite{Reik1967,Emin1993}, implying that this transition is subject to strong interactions with phonons. 

The central observation of this work is the appearance of strong asymmetric resonances at low temperatures in the infrared spectra of \TNS{} and \TNSe{} (shaded area in Fig.~1(b)). An expanded view of the optical conductivity (Figs.~ 2(a,c)) and permittivity (Figs.~2(b,d)) of \TNSe{} and \TNS{} shows that both compounds exhibit an unusual set of two resonances just above the optical absorption edge. The resonances are most clearly apparent as two distinct peaks in the permittivity spectra $\varepsilon_1(\omega)$ between 0.3 and 0.6 eV in \TNS{} and between 0.2 and 0.4 eV in \TNSe{} (black arrows in Figs.~2(b,d)). Figure 3 highlights the temperature evolution of the optical conductivity (Figs.~3(a,f)) and permittivity (Figs.~3(b,g)) in the vicinity of the resonances by displaying the difference spectra $\Delta \sigma_1 (\omega,T) = \sigma_1 (\omega, 10\ {\rm K})-\sigma_1(\omega, T)$ and $\Delta\varepsilon_1(\omega,T)=\varepsilon_1(\omega, 10\ {\rm K})-\varepsilon_1(\omega, T)$, respectively. The data show that the peaks arise in $\varepsilon_1(\omega)$ at low temperature, and that they are accompanied by strongly antisymmetric resonances in $\sigma_1(\omega)$. These features are distinct signatures of Fano interference between narrow discrete levels and a broad-band continuum \cite{Fano1961,Proepper2014}.

Before addressing the microscopic origin of the resonances, we discuss the quantitative description of the Fano interference, which is strongly constrained by the independent $\varepsilon_1(\omega)$ and  $\sigma_1(\omega)$ spectra extracted from our ellipsometric data. A detailed comparison of these data to the results of a Green's function analysis of the optical response of two narrow oscillators coupled to a continuum \cite{Zibold1992} yields the frequencies ($\omega_j$), spectral weights ($S_j$), and Fano asymmetry parameters ($q_j$) of the resonances (Appendix C). 
The total spectral weight of all infrared oscillators (comprising the resonances and the continuum) defines the effective number of electrons per Ni atom contributing to the associated transitions, $N_{IR}^\text{eff}=\frac{2m }{\pi e^2 N_\text{Ni}}\sum S_j$. The value obtained from our analysis, 0.34 (0.70) el/Ni for \TNS{} (\TNSe), is comparable to the density of the valence $\rm Ni$ $3d$ electrons. Figures 3(c,d,h,i) illustrate the strongly antisymmetric lineshapes of the narrow resonances near the absorption edges of \TNS{} and \TNSe{} (green and red lines and areas). In the vicinity of the resonance frequencies $\omega_{j=1,2}$ of 0.38 eV and 0.54 eV (0.23 eV and 0.33 eV) in \TNS{} (\TNSe), marked by vertical dotted lines in Fig.~3, the resonances can be fitted very well with a Fano line shape \cite{Fano1961} in the limit of strong coupling, with $q_{j=1,2}$ close to unity: 1.07 and 0.98 (1.29 and 1.28) for $j=1$ and $j=2$, respectively, implying comparable contributions of interfering discrete and continuum channels to the Fano resonances. Whereas the lineshapes of the Fano resonances in both compounds are closely similar, those in  \TNSe{} have much higher intensity. The effective number of electrons associated with the interference effects can be estimated by the spectral weight transferred across the resonance frequency, $N_{Fano}^\text{eff}\approx\frac{2m }{\pi
e^2 N_\text{Ni}}\int |\Delta\sigma_1 (\omega)| d \omega$, where $\Delta\sigma_1 (\omega)$ corresponds to the difference spectra $\sigma_1(10\,\text{K},\omega)-\sigma_1(T=150\,\text{K},\omega)$ in Figs.~3(a,f). Direct integration over the range of 0.1 to 0.7 eV yields 0.14 el/Ni for \TNSe, one order of magnitude higher than the corresponding value in \TNS{} ($\approx $ 0.01 el/Ni). Our detailed Green's function analysis, which allows for a more accurate determination of the resonance parameters, yields the same ratio of exciton peak intensities in \TNSe{} and in \TNS{} (Appendix C).
\begin{figure}
\includegraphics[width=0.5\textwidth]{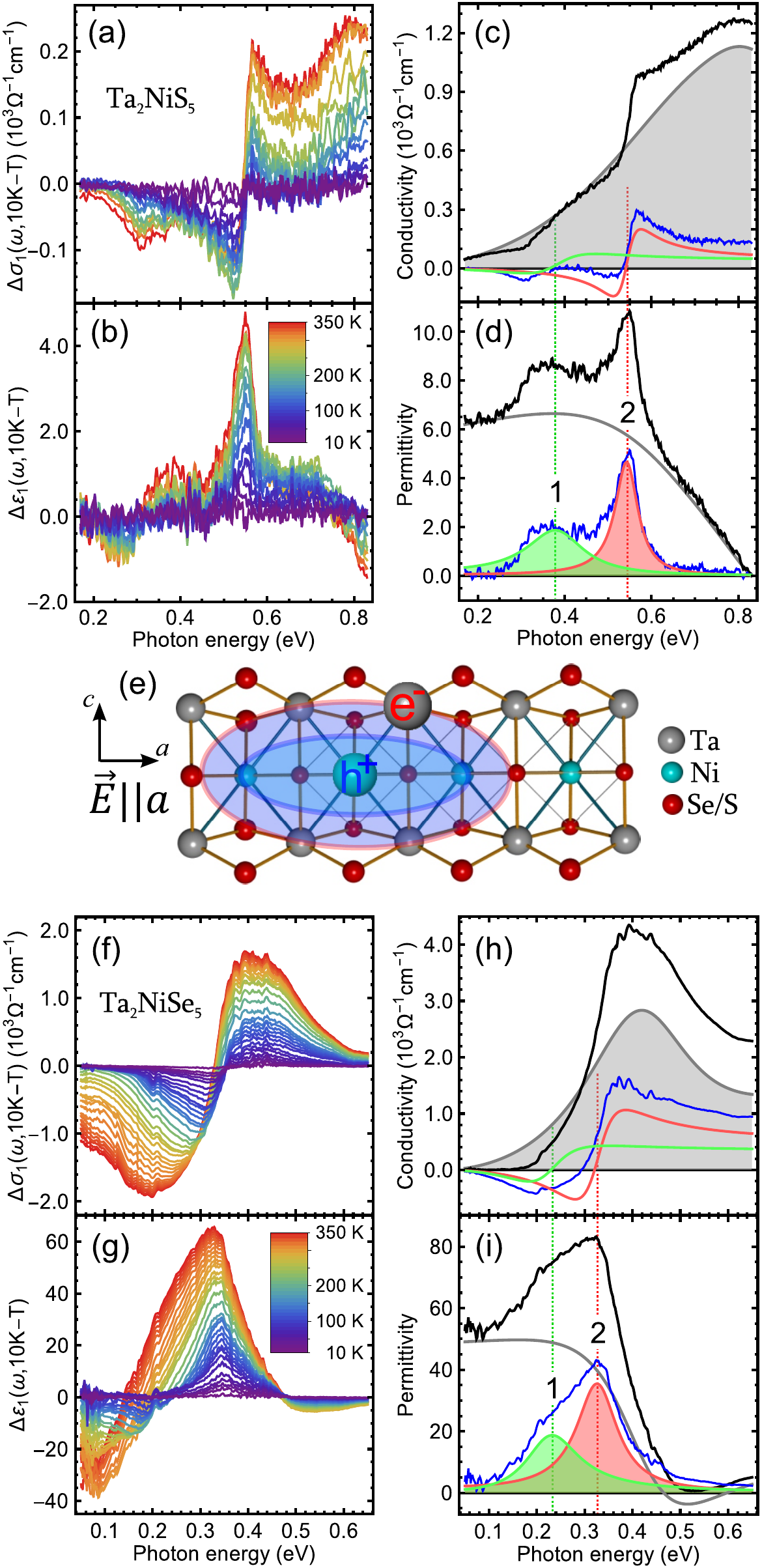}
\caption{
Difference spectra $\sigma_1(10\,\text{K},\omega)-\sigma_1(T,\omega)$ (a)
and $\varepsilon_1(10\,\text{K},\omega)-\varepsilon_1(T,\omega)$ (b) in \TNS.
(c,d) Real part of the optical conductivity $\sigma_1(\omega)$ (c) and dielectric
permittivity $\varepsilon_1(\omega)$ (d) of \TNS{} measured at $T=10$ K (black
lines). Red and green lines with shaded peaks represent the Fano fit to the
spectra (blue lines) obtained by subtraction of an electronic background
(shaded area and gray lines) from the data. The vertical dotted lines numbered
1 and 2 mark the resonance frequencies $\omega_{j=1,2}$. (e) Sketch of the
charge transfer exciton in the Ta-Ni chains along the $a$ axis. (f-i) are
the same as in (a-d) but for \TNSe. The gray and black lines in (d) are offset
by 16; cf. Fig.~2(b). 
}
\protect\label{fig3}
\end{figure}
\begin{figure}
\includegraphics[width=0.5\textwidth]{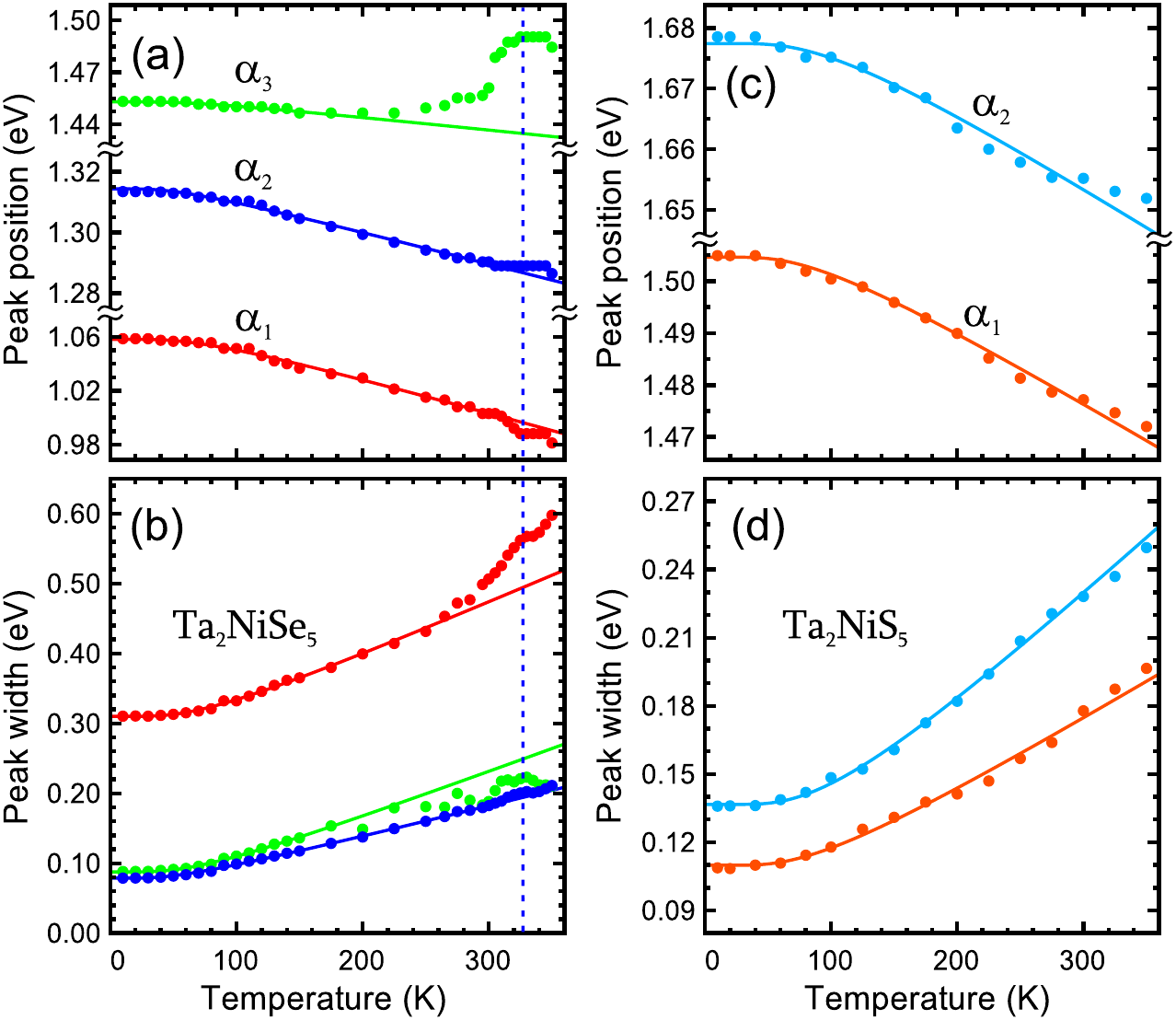}
\caption{
Temperature dependence of the peak energies and widths of the $\alpha$-absorption
bands (as in Fig.~1) for \TNSe{} (a, b) and \TNS{} (c, d). Solid lines are
fits proportional to the Bose-Einstein occupation number $n(\omega_{op}^j,
T)$ with the fitted parameters listed in Table I.}
\protect\label{fig4}
\end{figure}
In accord with prior theoretical work on \TNSe{} \cite{Kaneko2013}, we attribute the resonances to direct charge-transfer exciton states composed of holes in the hybridized Ni $3d$ and Se $4p$ \textquotedblleft molecular orbitals\textquotedblright{} and electrons in the $5d$ orbitals centered at four neighboring Ta ions, which form the valence and conduction bands, respectively. 
The resonance doublet can be viewed as part of an excitonic Rydberg series arising from the long-range Coulomb interaction, $E^{\text{ex}}_{n}$, with principal quantum numbers $n=1,2$. The energy splitting between the states in a Rydberg series is somewhat smaller than the exciton binding energy, $E_b=\alpha(\Delta E^{\text{ex}}_{2,1}), \ \alpha \gtrsim 1$, with $\Delta E^{\text{ex}}_{2,1}$ of 0.16 eV (0.1 eV) in \TNS{} (\TNSe). The parallel decrease of the exciton binding energy and the absorption edge energy with increasing size of the chalcogen atom is consistent with the notion that excitonic correlations drive the band gap. However, we expect the binding energy of excitons in the narrow Ni$-$Ta$-$Ni channels to be influenced not only by the Coulomb attraction between holes and electrons, but also by the degree of $d-p$ hybridization of the orbitals in which they reside \cite{Kaneko2013}. In the following we also argue that the excited states
self-localize via exciton dressing with a cloud of phonons. The associated
local deformation of the crystal lattice around the exciton can render the
Ta$-$Ni bonds in the chains oriented along $a$ axis (Fig.~3(e)) nonequivalent,
hence one can not exclude the possibility that the exciton splitting is caused
by lifting of the double degeneracy of the conduction band in the effective
three-chain model proposed in Ref. \cite{Kaneko2013}. Microscopic model calculations are therefore required before definitive conclusions about the nature of the ground state can be drawn based on the exciton energies.

We now turn to the most striking aspects of our experimental results, namely the unprecedentedly high spectral weight of the excitonic resonances and their strong Fano interference with continuum excitations. Excitonic Fano resonances, albeit with orders of magnitude smaller spectral weight, are well known in semiconductor physics, in situations where excitons in higher-energy sub-bands created by quantum confinement (and/or Landau or Wannier-Stark levels generated by external fields) are coupled to continuum states in lower sub-bands  \cite{Glutsch2013,BarAd1997}. Clearly, this mechanism cannot account for our observations on a bulk compound in the absence of external fields. However, Fano resonances can also arise if excitons are generated optically at energies above the continuum onset in bound states with either impurities or phonons. In view of the giant spectral weight of the resonances in  \TNSe{} (0.14 electrons per Ni atom), we rule out impurity bound states and focus on exciton-phonon bound states. In the following we will show that such states offer plausible explanations of the Fano interference, the giant spectral weight, and the temperature dependence of the resonance features in our spectra, as well as an intriguing perspective on the EI hypothesis.

The exciton-polaron binding mechanism is known to be highly effective in materials with quasi-1D electronic structure \cite{Ueta1986,Tokura2004}. In the compounds at hand, a hole centered at a Ni ion may be localized due to interactions with a breathing mode that affects the degree of $d-p$ hybridization such that a self-trapped excitonic bound state appears on the background of free electron-hole pairs. For \TNSe{} the real part of the far-infrared dielectric function corresponding to the total spectral weight of interband transitions, $\varepsilon_{tot}$, is extremely large due to the high polarizability of the $\rm Ta-Ni$ bonds ($\varepsilon_{tot} \approx 60$ for \TNSe{} in Fig.~2(d), compared to $\varepsilon_{tot} \approx 22$ for \TNS{} in Fig.~2(b)). The higher bond polarizability and lower characteristic energies make this compound very susceptible to structural distortions driven by the electron-phonon interaction and may stabilize the collective binding of electron-hole pairs in the system. 

Exciton-phonon interactions have indeed been proposed as an important ingredient of the EI transition and the monoclinic low-temperature structure of \TNSe{} \cite{Kaneko2013}. The minute monoclinic lattice distortion does not noticeably affect the interband optical transitions, but it may considerably weaken the interaction of excitons with the breathing phonon. Following this argument, we expect the binding energy of the exciton-polaron complexes in monoclinic \TNSe{} to be noticeably lower and the radius of trapped exciton wave function to be correspondingly larger than in \TNS.  As originally discussed by Rashba \cite{Rashba1975} in the context of excitons bound to isoelectronic impurities and/or optical phonons in conventional semiconductors, the formation of spatially extended bound states can lead to anomalous enhancement of the oscillator strength of the Fano resonances ("antenna effect").   In view of independent evidence of strong exciton-phonon coupling (see below), antenna emission of large exciton-polaron complexes is a highly plausible mechanism for the giant spectral weight of the resonances in the infrared spectra of \TNSe{} (Figs.~3(g,i)). Interestingly,  theoretical work on excitons bound to transition metal impurities in semiconductors \cite{Shibatani1968,Kikoin1994} shows that the Fano parameter $q$ is determined by the ratio of real
and imaginary lattice Green functions, weighted with the matrix elements
for impurity scattering and optical transitions, and thus unrelated to the oscillator strength. Similar considerations apply to exciton-phonon bound states as well, thus explaining the observation that the resonances in \TNS{} and \TNSe{} have similar $q$ while greatly differing in intensity.

Further evidence for strong exciton-phonon interactions comes from the analysis of the temperature dependence of the sharp absorption bands at energies above the excitonic Fano resonances ($\alpha$ through $\delta$ in Fig.~1), and from the rapid evolution of the peak parameters near the structural transition temperature in \TNSe{} (Fig.~4). As an example, Figs. 4(a,b) [(c,d)] present the temperature dependence of the peak parameters of the $\alpha$-absorption bands in \TNSe{} [\TNS], which were extracted from fits to symmetric Lorentz profiles (Appendix C). The energy $\omega^j$ and width $\Gamma^j$ of each oscillator $j$ were then simultaneously fitted to the expressions %
\begin{eqnarray}
\omega_j(T) = \omega_0^j-\omega_1^j n(\omega_{op}^j, T), 
\nonumber
\\*
\Gamma_j(T) = \Gamma_0^j+\Gamma_1^j n(\omega_{op}^j, T), 
\end{eqnarray}
where $n(\omega_{op}^j, T)^{-1}=\exp{\frac{\hbar\omega_{op}^j}{k_B T}-1}$  is the Bose-Einstein occupation number of a phonon mode with frequency $\omega_{op}$. The fitted parameters are presented in Table I. The excellent agreement of these expressions with the experimental data (Fig.~4(c-f)) implies the coupling of an optical phonon to the photoexcited states, closely analogous to the thermal broadening of charge-transfer bands in halogen-bridged nickel chain compounds, which was attributed to exciton-optical phonon interactions \cite{Tokura2004}. The phonon energies resulting from the fits, $\omega_{op}$ of $21\div 24$ meV for \TNS{} and $13\div 17$ meV for \TNSe, agree well with the energy range of optical phonons in these materials and with the Se/S mass ratio $\omega^{\rm S}_{op}/\omega^{\rm Se}_{op}\approx \sqrt{m^{\rm Se}/m^{\rm  S}}$ of 1.57.  At the structural phase transition in \TNSe, the measured peak parameters undergo rapid changes analogous to those of the low-energy optical excitations (Fig.~2), thus confirming the key role of electron-phonon interactions in driving this transition.
\begin{table}[t]
\centering
\caption{The fitted parameters for the temperature
dependences of the $\alpha$-absorption bands in \TNSe{} and \TNS{} according
to Eqs. (1).}
    \begin{tabular}{ccccc}
        \multicolumn{5}{c}{\TNSe{}}                \\
        \hline
        $\omega_0\text{ (eV)}$    & $\omega_1\text{ (meV)}$    &
        $\Gamma_0\text{ (meV)}$    & $\Gamma_1\text{ (meV)}$    &
        $\omega_{\text{op}}\text{ (meV)}$        \\
        \hline                       
        1.45 &  8.2 &  87.5 &  89.3 & 13.7 \\
        1.31 & 15.5 &  78.5 &  65.5 & 12.7 \\
        1.06 & 50.8 & 309.9 & 143.5 & 16.6 \\
        \\
        \multicolumn{5}{c}{\TNS{}}                \\
        \hline
        $\omega_0\text{ (eV)}$    & $\omega_1\text{ (meV)}$    &
        $\Gamma_0\text{ (meV)}$    & $\Gamma_1\text{ (meV)}$    &
        $\omega_{\text{op}}\text{ (meV)}$        \\
        \hline                       
        1.67 & 30.2 & 135.9 & 121.8 & 21.8 \\
        1.50 & 39.1 & 109.6 &  96.7 & 23.1
    \end{tabular} 
\end{table} 

\section{Summary}
In summary, our ellipsometric experiments on \TNSe{} and \TNS{} have uncovered an unusual set of excitonic resonances that are generated optically in self-trapped states due to strong interaction with phonons. The giant spectral weight of the excitonic states in \TNSe{} indicates that they are highly extended and strongly overlapping along the Ta$-$Ni chain direction. These unusual conditions are highly conducive for the formation of an excitonic condensate, in accord with the prediction that the low-temperature phase of \TNSe{} should be regarded as an excitonic insulator. Although our experiments do not provide direct access to the ground state, the resulting wealth of information on excitons and their many-body interactions now enable detailed, microscopic investigations of the excitonic insulator hypothesis.

\begin{acknowledgments}
We gratefully acknowledge T.~M.~Rice, I.~I.~Mazin, and F.~V.~Kusmartsev
for fruitful discussions.
\end{acknowledgments}

\appendix
\section{Anisotropy correction for \TNSe{}}
\begin{figure}
\includegraphics[width=0.5\textwidth]{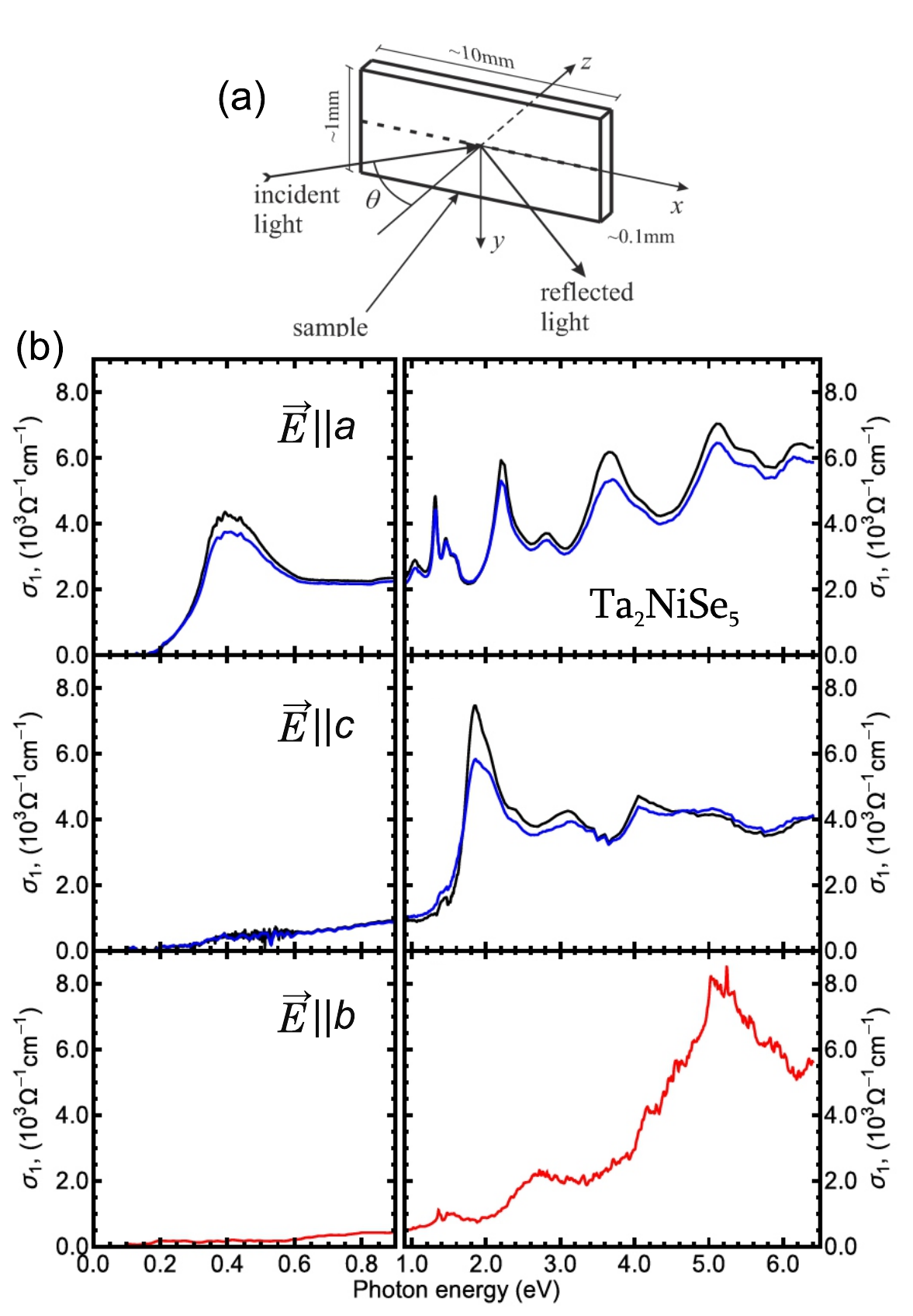}
\caption{
(a) Sketch illustrating the frame of reference and dimensions of the sample.
$\theta$ is the angle of incidence. (b) Real part of the optical conductivity
$\sigma_{1}$ for polarizations along the principal axes of \TNSe{}. The black
curves are obtained from the as-measured pseudodielectric functions according
to Eq.(5), the blue curves are the true response fitted as detailed in the
text, and the red curve is the calculated $\sigma_{1yy}(\omega)$ (the same
as in Fig.~S1(a)) conductivity spectrum used as an approximation for fitting
the true values of the optical conductivity.}
\protect\label{fig5}
\end{figure}
For an anisotropic
material with a dielectric tensor of the form
\begin{equation}
\boldsymbol{\hat{\varepsilon}} =       
\begin{pmatrix}
                                  \varepsilon_x &             0 & 0     
         \\
                                              0 & \varepsilon_y & 0     
         \\
                                              0 &             0 & \varepsilon_z
                
\end{pmatrix}
\end{equation}
 the complex reflectance ratio $\rho(\varepsilon_x,\varepsilon_y,\varepsilon_z,\theta)
= \tilde{r_p} / \tilde{r_s}$ is given by the expression
\begin{equation}
                        \rho = \frac{\cos \theta \sqrt{\varepsilon_x} - \sqrt{1
- \frac{\sin^2 \theta}{\varepsilon_z}}}{\cos \theta \sqrt{\varepsilon_x}
+ \sqrt{1 - \frac{\sin^2 \theta}{\varepsilon_z}}} \cdot \frac{\cos \theta
+ \sqrt{\varepsilon_y - \sin^2 \theta}}{\cos \theta - \sqrt{\varepsilon_y
- \sin^2 \theta}},
\end{equation}
where the frame of reference is defined as shown in Fig.~5(a).
The measurements were first done as shown ($a \parallel x,$ $c \parallel
y$) then the sample was rotated 90$^{\circ}$ around the $z$-axis.
To obtain the true values of the dielectric tensor we take the measured values
of $\rho_x$ and $\rho_y$ and simultaneously fit $\varepsilon_x$ and $\varepsilon_y$
to satisfy the equations
\begin{equation}
\rho_x = \rho(\varepsilon_x,\varepsilon_y,\varepsilon_z,\theta),
\
\rho_y = \rho(\varepsilon_y,\varepsilon_x,\varepsilon_z,\theta),
\end{equation}
while using the calculated spectra for $\varepsilon_z$ (Appendix B, Fig.~7(a)). In the
case of our samples, $\rho_x$ would correspond to value of $\rho$ measured
when the $a$-axis of the sample is oriented along $x$ and the $c$-axis along
$z$, as shown in Fig.~5(a), and vice versa for $\rho_y$. Such a fit would
then yield $\varepsilon_x=\varepsilon_a$ and $\varepsilon_y=\varepsilon_c$
(i.e. $\varepsilon_{a,c}$ being the dielectric response of the material for
$E \parallel a,c$). The results of this fit for \TNSe{} are shown in Fig.~5(b).
along with the analytically inverted function
\begin{equation}\label{PseudoEpsilon}
                        \tilde{\varepsilon}_{a,c} = \tan^2 \theta \frac{1
+ 2 \rho_{x,y}
\cos (2 \theta) + \rho_{x,y}^2}{(1 + \rho_{x,y})^2},
\end{equation}
which is referred to as the pseudodielectric function. If the material is
isotropic and non-magnetic, it is equal to the true dielectric function.
Figure 3 shows that the corrected-for-anisotropy dielectric function is
not considerably different from the measured pseudodielectric function, retaining
the shape and amplitude of most features unchanged.
Repeating the measurements at various angles of incidence $\theta$ helps
one to improve the anisotropy correction. 
The reported $a$-axis component of $\varepsilon (\omega)$ corresponds to
the measured pseudodielectric function $\langle \varepsilon_a\rangle
\approx \varepsilon_a$ at angle of incidence ranged from 70$^{\circ}$ to
80$^{\circ}$ for sample
orientations with the $a$ axis in the plane of incidence.

\begin{figure}
\includegraphics[width=0.5\textwidth]{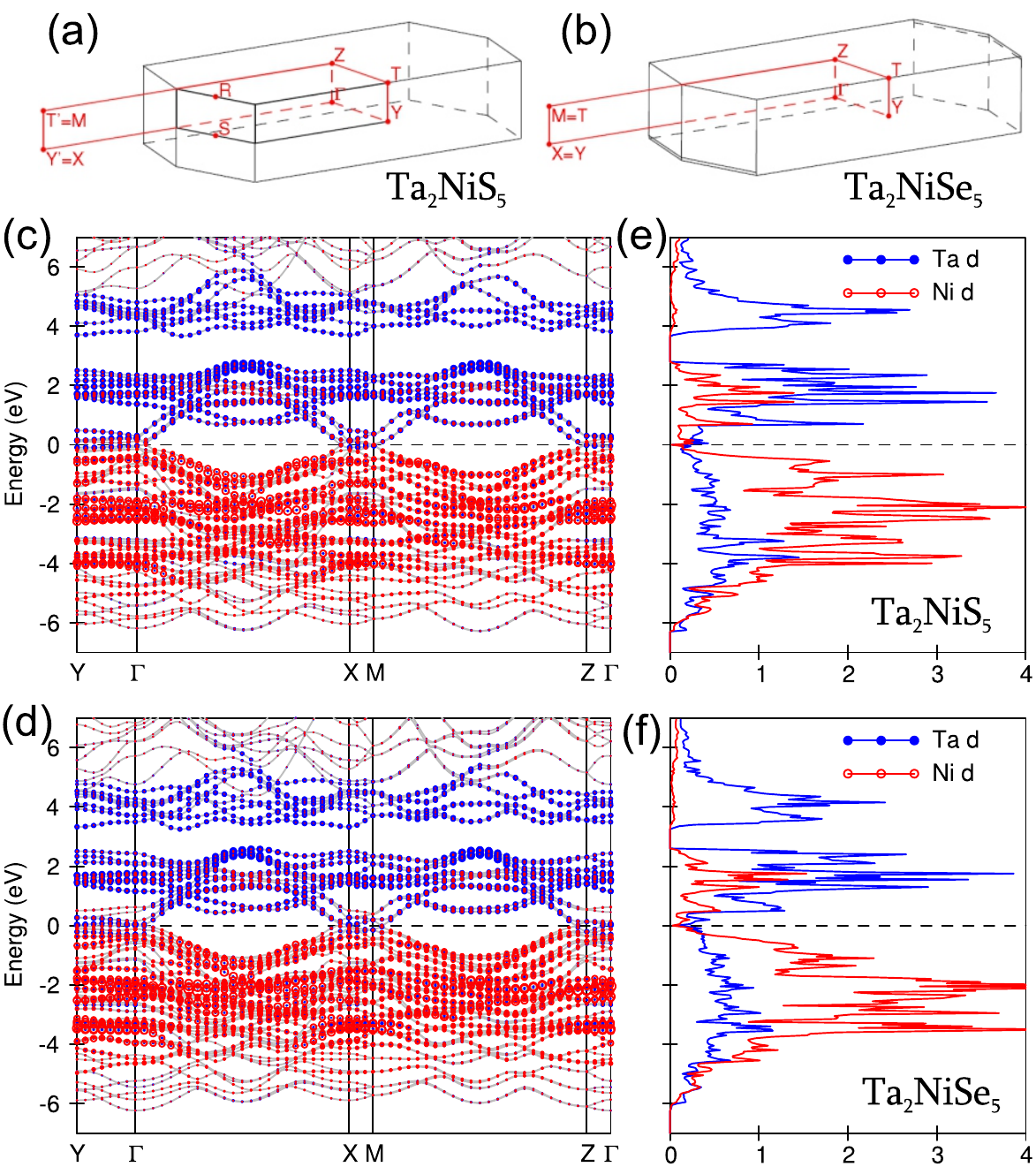}
\caption{
(a, b) Brillouin zone of {\it Cmcm} \TNS{} (\textbf{a}) and {\it C2/c} \TNSe{}
(\textbf{b}). (c-f) Relativistic band structure and
density of states projected on the Ta $5d$ (blue)  and Ni $3d$ (red) atomic
states of \TNS{} (c, e) and \TNSe{} (d, f).}
\protect\label{fig6}
\end{figure}

\section{Electronic structure calculations}
Figure 6 shows the result of the density functional theory (DFT) calculations
using the linear muffin-tin orbital (LMTO) method within the atomic sphere
approximation \cite{Yaresko2004} based on the experimental crystal structures
of \TNS{} and \TNSe{} \cite{Sunshine1985}. Although the symmetry of the \TNSe{}
crystal structure lowers to monoclinic, the band structures calculated for
both compounds are very similar. Accounting for on-site Coulomb repulsion is not sufficient to open a gap within the LDA+U band-structure approach, while ARPES, dc transport, and optical measurements indicate a gap. Kaneko {\it et al.} \cite{Kaneko2013} remedy the discrepancy by applying orbital-dependent potentials to the calculation, beyond a conventional {\it ab initio} approach.
 
The DFT calculations reproduce only the general trend of the material and
polarization dependences of the band structure and optical conductivity $\sigma_1(\omega)$
(Fig.~7). The Ta 5$d$ band dispersion along the chain direction (i.e. along
the $\rm \Gamma - X$ line in the Brillouin zone in Fig.~6(a,b)) is clearly stronger than the band dispersion along two other directions in Figs.~6(c,d). The shorter  Ta$-$S bonds and stronger Ta $d$ -- S $p$ hybridization as compared
to Ta$-$Se ones result in the blue shift of the corresponding interband transitions in $\sigma_1(\omega)$ in Fig.~7.
\begin{figure}  
\includegraphics[width=0.5\textwidth]{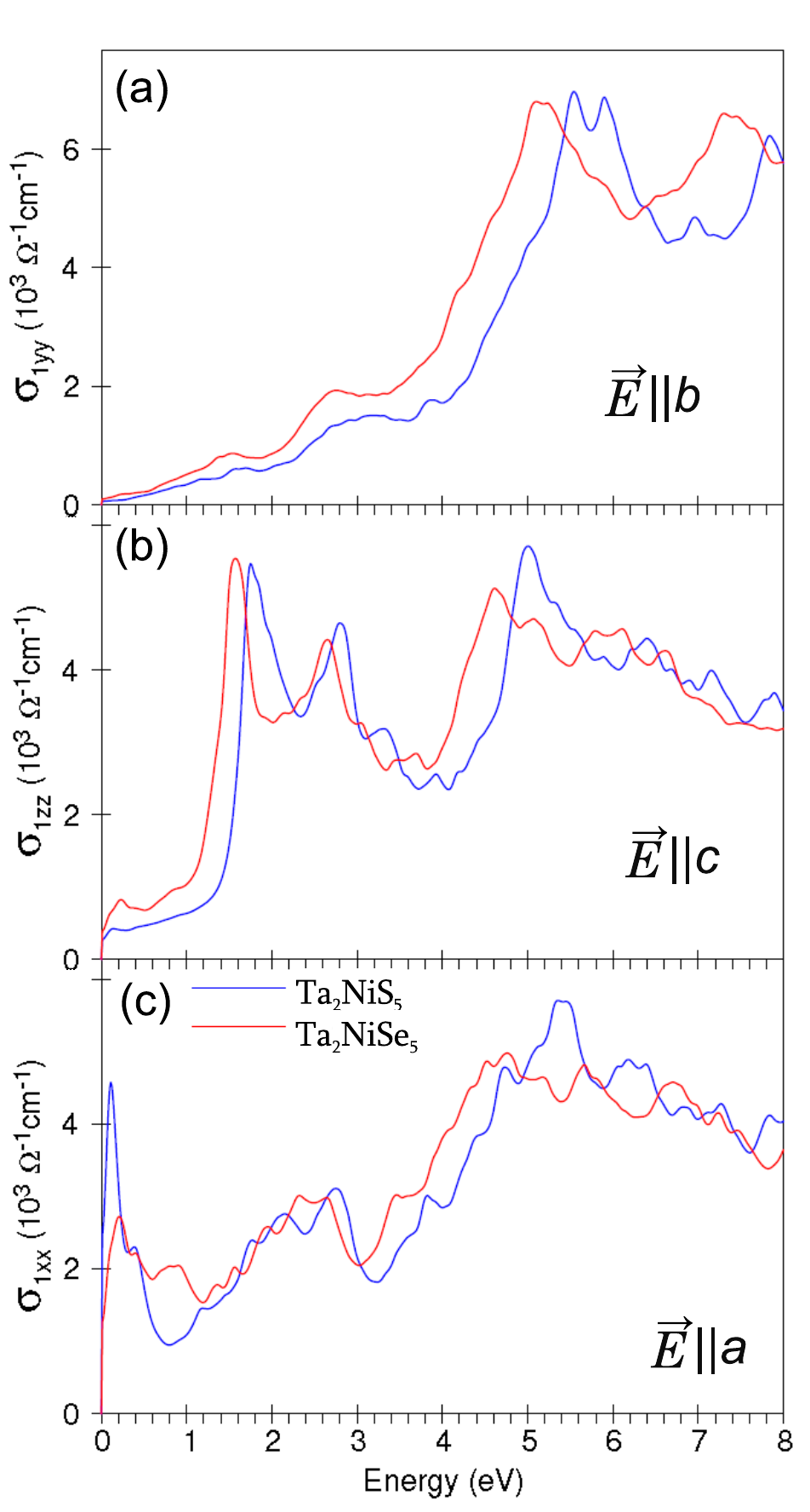}
\caption{
Calculated diagonal elements of the $\sigma_{1yy}(\omega))$ (a),
$\sigma_{1zz}(\omega))$ (b), and $\sigma_{1xx}(\omega))$ (c) optical conductivity
of \TNS{} (blue) and \TNSe{} (red).
The Drude contribution to the optical conductivity due to excitations within
the bands crossing $E_F$ are not included in the spectra.}
\protect\label{fig7}
\end{figure}
\section{Green's function analysis of the Fano resonances}

In order to quantitatively parameterize the observed exciton resonances in
\TNSe{} and \TNS, we model the complex dielectric function $\varepsilon(\omega)=\varepsilon_1(\omega)+4\pi i
\sigma_1 (\omega)/\omega$ of a coupled system of two narrow oscillators and
a single broad continuum by applying the phenomenological approach based
on the Green's function formalism \cite{Zibold1992,Burlakov1992,Loidl2004,BarAd1997,Glutsch2013}.
 
We  fit the following expression simultaneously to the experimentally measured
$\sigma_1(\omega)$ and $\varepsilon_{1}(\omega)$ spectra:
\begin{equation}
        \varepsilon_1(\omega)+4\pi i\sigma_1 (\omega)/\omega
 = \bold{A}\hat{\mathrm{G}} \bold{A} + \tilde\varepsilon_{\textsc{bg}}(\omega),
\end{equation}
where $\bold{A}=(A_e,A_1,A_2)$ is a vector of matrix elements for dipole
transitions, $S_{e}=A_{e}^2/8$ and $S_{j=1,2}=A_{j=1,2}^2/8$ are the spectral
weights of the broad electronic oscillator and of the $j$th exciton, respectively,
and
$\hat{\mathrm{G}}$ is the Green's function matrix
\begin{equation}
                \hat{\mathrm{G}}^{-1} = \begin{pmatrix}
                 G_e^{-1} & i k_1\omega & i k_2\omega \\
                i k_1\omega & G_1^{-1}& 0 \\
                i k_2\omega & 0 & G_2^{-1} 
        \end{pmatrix}.
\end{equation}\
The Green's functions are $G_{e,1,2}^{-1} = \omega_{e,1,2}^2 - i \gamma_{e,1,2}
\omega - \omega^2$,
where $\omega_{j=1,2}$ and $\gamma_{j=1,2}$ are the resonance frequency and
the damping constant of the $j$th exciton , $\omega_e$ and $\gamma_e$ are
the resonance frequency and the damping constant of the electronic oscillator,
respectively, and $i k_{j=1,2}\ \omega$ is the coupling constant
for interaction of the $j$th exciton with the continuum. 

We separate the contribution of the coupled oscillators
$\tilde\varepsilon_{\textsc{IR}}(\omega)$
from the background spectrum $\tilde\varepsilon_{\textsc{bg}}(\omega)$ of
the higher-energy non-interacting oscillators labeled $\alpha$ through $\delta$
(Fig.~1 of the manuscript) by subtracting a sum of classical symmetric Lorentz
oscillators of the fit to $\sigma_1(\omega)$ and $\varepsilon_{1}(\omega)$
with resonance energies approximately equal to, or higher than 1~eV:
\begin{equation}\label{eq:GLorentz}
 \tilde\varepsilon_{\textsc{bg}}(\omega)  
   =\varepsilon_\infty+\sum^{\alpha,\beta,\gamma,\delta}_{j=1}\frac{\Delta\varepsilon_j\omega^2_{j}}{\omega^2_{j}-\omega^2-\mathrm{i}\omega\gamma_j},
\end{equation}
where $\omega_j$ and $\gamma_j$ are the resonance frequency and the damping
of the $j$th oscillator, respectively, $\Delta\varepsilon_j$ is its contribution
to the static permittivity, and $\varepsilon_\infty$ represents the contribution
to the dielectric function by transitions above the experimentally accessible
energy range $>$ 6.5~eV. 

With the exception of $\tilde\varepsilon_{\textsc{bg}}(\omega)$ the expansion of Eq. (C1) reads \begin{eqnarray}
\tilde\varepsilon_{\textsc{ir}}(\omega)=\varepsilon(\omega) - \tilde\varepsilon_{\textsc{bg}}(\omega)=
\nonumber
\\
= \sum_{j=1,2} A_j^2G_j + \frac{(A_e - \sum_{j=1,2} i k_{j}\omega
G_j A_j)^2}{G_e^{-1} + \sum_{j=1,2} k_{j}^2\omega^2G_j},
\end{eqnarray}
where $A_{e,j}$, $\omega_{e,j}$, $\gamma_{e,j}$, and $k_j$ are used as independent
fit parameters and listed in Table II. The corresponding infrared dielectric
functions are presented in Fig.~3(c,d) for \TNS{} and Fig.~3(h,i) for \TNSe{}
of the manuscript. By choosing the coupling constant to be imaginary ($i k_{j}\omega$),
we achieve a solution close to that by Bad-Ar {\it et al.} \cite{BarAd1997}.
The authors of Ref. \cite{BarAd1997} take a quantum-mechanical approach and solve the Dyson equation to calculate the response of a discrete state to a quasi-continuum
of closely spaced discrete energy levels in the Fano-Anderson model. There
the parameters of the unperturbed Green's functions of the discrete transitions
(equivalent to $G^{-1}_{j=1,2}$ in Eq. (C2)) are given directly as the intrinsic
({\it "bare" or "unperturbed"}) parameters of the discrete state.
\begin{table}
        \caption{Green's function model parameters for the infrared spectra
of \TNSe{} and \TNS{} at 10~K, for Eq. (C4). The spectral weights $N^\text{eff}_j$
are given in electrons per Ni atom, $N^\text{eff}_j=\frac{2m}{\pi e^2 N_\text{Ni}}S_j\
(S_j=A_j^2/8)$.}
        \centering
    \begin{tabular}{c | r@{.}l r@{.}l r@{.}l r@{.}l r@{.}l}
                            \multicolumn{11}{c}{\TNSe} \\ \hline
                            $j$ & \multicolumn{2}{l}{$N^\text{eff}_j \ \
\ \ \  $} 
                                                                & \multicolumn{2}{c}{$A_j$
(eV)} 
                                                                & \multicolumn{2}{c}{$\omega_j$
(eV)}
                                                                & \multicolumn{2}{c}{$\gamma_{j}$
(eV)} 
                                                                & \multicolumn{2}{c}{$k_{j}$
(eV)} \\ 
                                                                \hline
                                                        $e$ & 0&643 &  2&248
& 0&543 & 0&302 & \multicolumn{2}{c}{---} \\
                \textcolor{green}{1}    & 0&035 &  0&524 & \textcolor{green}{0}&\textcolor{green}{209}
& 0&110 & 0&230 \\
                \textcolor{red}{2}              & 0&019 &  0&381 & \textcolor{red}{0}
 &\textcolor{red}{312}   & 0&111 & 0&106 \\ \multicolumn{11}{c}{} \\
                
                                                              \multicolumn{11}{c}{\TNS}
\\ \hline
                            $j$ & \multicolumn{2}{l}{$N^\text{eff}_j \ \
\ \ \  $} 
                                                                & \multicolumn{2}{c}{$A_j$
(eV)} 
                                                                & \multicolumn{2}{c}{$\omega_j$
(eV)}
                                                                & \multicolumn{2}{c}{$\gamma_{j}$
(eV)} 
                                                                & \multicolumn{2}{c}{$k_{j}$
(eV)} \\ 
                                                                \hline
                                                        $e$ & 0&335 &  1&715
& 0&703 & 0&531 & \multicolumn{2}{c}{---} \\
                \textcolor{green}{1}    & 0&001 &  0&103 & \textcolor{green}{0}&\textcolor{green}{371}
& 0&126 & 0&113 \\
                \textcolor{red}{2}              & 0&003 & -0&172 & \textcolor{red}{0}
 &\textcolor{red}{526}   & 0&099 & 0&164 \\
        \end{tabular}
\end{table} 
 
                         \begin{table}
                         \caption{Generalized Lorentz oscillators' parameters
 for the infrared spectra of \TNSe{} and \TNS{} at 10~K, for Eq. (C5).}
                                 \centering
                                 \begin{tabular}{c | r@{.}l r@{.}l r@{.}l
 r@{.}l | r@{.}l r@{.}l r@{.}l r@{.}l}
                                           & \multicolumn{8}{c|}{\TNSe{}}
        &
 \multicolumn{8}{c}{\TNS{}}              \\ \hline
                                         $j$ & \multicolumn{2}{c}{$S'_j$, $\mathrm{eV}^2$}
         & \multicolumn{2}{c}{$\omega'_j$, eV} & 
                                               \multicolumn{2}{c}{$\gamma'_j$,
 eV}                     & \multicolumn{2}{c|}{$\beta_j$, eV} & 
                                               \multicolumn{2}{c}{$S'_j$, $\mathrm{eV}^2$}
         & \multicolumn{2}{c}{$\omega'_j$, eV} & 
                                               \multicolumn{2}{c}{$\gamma'_j$,
 eV}                     & \multicolumn{2}{c}{$\beta_j$, eV} \\ \hline
                                         $e$ &     {} 3&828 & {} 0&450 & {}
 0&285 & -6&789 &  2&769 & {} 0&672 & {} 0&571 & -0&443  \\
                                         \textcolor{green}{1} & {} 0&154 &
 {} \textcolor{green}{0}&\textcolor{green}{238} & {} 0&129
 &  2&475 & 0&004 & {} \textcolor{green}{0}&\textcolor{green}{378} & {} 
 &  0&164                \\
                                         \textcolor{red}{2} & {} 0&356 & {}
 \textcolor{red}{0}&\textcolor{red}{332} & {} 0&111
&  4&314 &  -0&004 & {} \textcolor{red}{0}&\textcolor{red}{542} & {} 0&062
&  0&279                \\
                               \end{tabular}
                   \end{table}

                        \begin{table}
                   \caption{Strength $p_{j}^\text{eff}$ (Eq. (C8)) and asymmetry
 $q_j$ (Eq. (C7)) parameters of the exciton Fano resonances in \TNSe{} and
\TNS{} at 10~K, given the values from Table III.}    
                        
                                \centering
                                \begin{tabular}{c | r@{.}l r@{.}l | r@{.}l
r@{.}l}
                                & \multicolumn{4}{c|}{\TNSe{}} & \multicolumn{4}{c}{\TNS{}}
\\ \hline
                                        $j$ & \multicolumn{2}{l}{$p_{j}^\text{eff}$
($el/{\rm Ni}$)}
   &
\multicolumn{2}{c|}{$q_j$} & 
                                              \multicolumn{2}{l}{$p_{j}^\text{eff}$
($el/{\rm Ni}$)}
   &
\multicolumn{2}{c}{$q_j$}       \\ \hline
                                      
                                        \textcolor{green}{1} & {} {} 0&078
&  1&294 & {} {}
0&007 &   1&066 {} \\
                                        \textcolor{red}{2} & {} {} 0&188
&  1&279 & {} {}
0&017 &   0&976 {} \\
                                \end{tabular}
                        \end{table}
                        
Expanding the expression for $\tilde\varepsilon_{\textsc{IR}}(\omega)$ in
Eq.(C4) gives a rational function, which is equivalently represented as
\begin{equation}
        \tilde\varepsilon_{\textsc{ir}}(\omega) =  \sum_{j=e,1,2} \frac{S'_j
- i \beta_j \omega}{\omega'^2_j - i \gamma_j' \omega
- \omega^2},
\end{equation}
i.e. as a sum of generalized Lorentzian oscillators, augmented by an asymmetry
parameter ($\beta_j$). Because $\tilde\varepsilon_{\textsc{ir}}(\omega)$
in Eq. (C4) decays as fast as $\omega^{-2}$, the condition $\sum_{j=e,1,2}\beta_j=0$
holds, satisfying the causality of the dielectric response. 
This allows to fit a set of generalized Lorentzian oscillators simultaneously
to the experimentally measured $\sigma_1(\omega)$ and $\varepsilon_{1}(\omega)$
spectra over the entire spectral range, as in a number of other studies,
e.g. in Refs. \cite{Humlichek2000,Proepper2014}. Generalized oscillators
with $\beta_j=0$ (equivalent to classical symmetric Lorentzians in Eq.(C3)
with $S'_j=\Delta\varepsilon_j\omega^2_{j}$) were sufficient to describe
most of the interband transitions ($\alpha$ through $\delta$ sets in Fig.~1
of the manuscript) in the measured spectra (assigned as $\tilde\varepsilon_{\textsc{bg}}(\omega)$
above) with only three resonances carrying asymmetric shapes, which we associate
with the coupled oscillators and fit with the Green's function matrix equation
Eq.(C1). The green and red lines and areas in Fig.~3(c,d) for \TNS{} and Fig.~3(h,i) for \TNSe{} in the
manuscript correspond to $\tilde\varepsilon_{j=1,2}$, respectively, while
the gray lines and areas are given by $\tilde\varepsilon_e + \tilde\varepsilon_\textsc{bg}$.
The obtained parameters $S'_j$, $\omega'_j$, $\gamma'_j$, and $\beta_j$ for
the asymmetric oscillators are listed in Table III.  The requirement $\sum_{j=e,1,2}\beta_j=0$
is satisfied, $\beta_1+\beta_2=-\beta_e$. Since the broad electronic band
has much larger spectral weight and width, its asymmetry is not evident.

In the vicinity of the resonance frequency the generalized Lorentz oscillator
can be converted into the Fano profile resulting from the quantum interference
of a discrete state with a continuum \cite{Fano1961}:
\begin{equation}\label{eq:Fano}
\Delta\sigma^j_1(\omega)=\sigma^j_0\frac{(q_j+\epsilon)^2}{1+\epsilon^2}\,,
\end{equation}
where 
\begin{equation}
        \epsilon=2(\omega-\omega'_j)/\gamma'_j \text{ and } q_j \approx \frac{S'_j
+ \sqrt{S'^2_j + \beta_j^2 \omega'^2_j}}{\beta_j \omega'_j}.
\end{equation}
The Fano parameters $q_j$ are listed in Table IV along with the strength
parameter $p_{j}^\text{eff}$, which provides a measure of the spectral weight
of the $j$th narrow generalized Lorentzian,
\begin{equation}
p_{j=1,2}^\text{eff}\approx\frac{2m}{\pi e^2 N_\text{Ni}}
\frac{\sqrt{S'^2_j + \beta_j^2 \omega'^2_j}}{8},
\end{equation}
and defines the effective number of electrons per Ni atom contributing to
the associated interference effects.


The parameters of the Green's functions according to Eq. (C4) are given as
the intrinsic {\it"unperturbed"} parameters of the discrete states, whereas
the {\it"perturbed"} parameters in Eq. (C5) do reflect the positions and
shapes of the features in the spectra. Acknowledging the phenomenological
character of the applied models, we do not discuss the microscopic origin
of the {\it"unperturbed"} oscillators' parameters and their relation to the
coupling strength. Nevertheless, one can, in particular, estimate how much
spectral weight is transferred from the background continuum to the discrete
states, by comparing the values $N^\text{eff}_1+N^\text{eff}_2=0.054\ el/{\rm
Ni}$ $(0.004\ el/{\rm Ni})$ and $p_1^\text{eff}+p_2^\text{eff}=0.266\
el/{\rm Ni}$ $(0.024\ el/{\rm Ni})$ for \TNSe{} (\TNS) according to Tables
II and IV, respectively. Though the total weight of the spectrum perturbed
by the interference effects exceeds the intensity of the {\it"unperturbed"}
exciton states by a factor of $\sim 5$, 0.054 electrons per Ni atom in \TNSe{}
is still uncharacteristically large for excitonic peaks. Regardless of the
model used, the exciton peak intensities in \TNSe{} are order of magnitude
larger than those in \TNS.

\bibliography{/Users/boris/Documents/TodayPaper/CondMatsubm/LarkinFanoArxivFeb2017}

\end{document}